\documentclass[aps,pre,twocolumn]{revtex4-2}
\usepackage{graphicx}
\usepackage[dvipdfmx]{hyperref}

\newcommand{\be}{\begin{eqnarray}}
\newcommand{\ee}{\end{eqnarray}}

\begin{document}

\title{Dynamical Lee--Yang zeros for continuous-time and 
discrete-time stochastic processes}
\author{Hiroki Yoshida}
\affiliation{Department of Physics, Tokyo Institute of Technology,
Tokyo 152--8551, Japan}
\author{Kazutaka Takahashi}
\affiliation{Institute of Innovative Research, Tokyo Institute of Technology,
Kanagawa 226--8503, Japan}

\date{\today}

\begin{abstract}
We describe classical stochastic processes
by using dynamical Lee--Yang zeros.
The system is in contact with external leads and the time evolution 
is described by the two-state classical master equation.
The cumulant generating function is written in a factorized form
and the current distribution is characterized
by the dynamical Lee--Yang zeros.
We show that a continuous distribution of zeros is obtained 
by discretizing the time variable.
When the transition probability is a periodically oscillating function 
of time, the distribution of zeros splits into many parts.
We study the geometric property of the current 
by comparing the result with that of the adiabatic approximation.
We also use the Floquet--Magnus expansion in the continuous-time case 
to study dynamical effects on the current at the fast-driving regime.
\end{abstract}
   
\maketitle

\section{Introduction}
\label{introduction}

Understanding statistical fluctuations of current
is one of the main objectives in nonequilibrium physics.
When the system is coupled to multiple reservoirs,
we observe a current through the system.
The fluctuation theorem tells us
how the underlying symmetry of the system is reflected to
the current distribution~\cite{ECM, GC, Crooks}.

In the long-time limit,
the system settles down to a stationary behavior
irrespective of the choice of the initial condition.
Full counting statistics describes distributions of transferred charges
through the system~\cite{GC, LL, LLL, BN, SU}.
The generating function can be treated as
an analog of the partition function
in equilibrium statistical mechanics.

The analytic properties of the partition function can be
described by the Lee--Yang zeros~\cite{YL, LY, Fisher}.
The partition function is a positive quantity and
goes to zero only when we set the parameters to unphysical values.
The only exception is when the system has a phase transition.
It involves a breakdown of the analyticity
and can be described by the distribution of the zeros.

A similar behavior is also expected to hold in
the generating function of the full counting statistics.
Previous studies found that the generating function is described 
by the dynamical Lee--Yang 
zeros~\cite{AI, IA, FG, UEUA, HFG, SMY16, BMPGF, SMY17}.
We note that
the Lee--Yang zeros are useful not only 
in the presence of the phase transitions 
but also for general systems.
By knowing the distribution of zeros, 
we can obtain the complete information of the system.
Any statistical quantities are represented by zeros.
It was shown in quantum systems~\cite{WL, PZWCDL}
and in classical stochastic systems~\cite{BMPGF}
that the Lee--Yang zeros are not artificial mathematical objects 
but are directly related to the experimental observables.

Zeros of the generating function can be clearly understood
when the number of events is finite, or countably infinite.
For a stochastic process described by
the probability distribution $\{P_n\}_{n=-M,-M+1,\dots,M}$,
the generating function is defined as
\be
 Z(\chi)=\sum_{n=-M}^M (e^\chi)^nP_n.
\ee
When the positive integer parameter $M$ is finite, 
the generating function, written as 
\be
 Z(\chi)=\frac{1}{(e^{\chi})^{M}}
 \prod_{i=1}^{2M} \frac{e^{\chi}-z_i}{1-z_i}, \label{Zpoly}
\ee
is completely characterized
by a set of zeros $\{z_i\}_{i=1,\dots,2M}$.
Each $z_i$ represents
a point where the generating function goes to zero
in complex plane of $z=e^\chi$.
Statistical quantities calculated from the generating function
are represented by using the zeros.

The number of zeros becomes large as the number of events increases.
However, previous studies found that
the number of zeros in stochastic systems described by the
two-state classical master equation
is given by a small finite value~\cite{IA, FG, HFG, BMPGF}.
This property is totally different from the original Lee--Yang theory where
many zeros are accumulated to form a continuous distribution in
a complex plane in the thermodynamic limit.
To find a similar behavior in stochastic systems,
we need to find a controllable parameter that determines the number of zeros.

A different interesting problem is the situation 
when the system is driven periodically.
Then, the transition-rate matrix in the master equation is time dependent.
When we treat the system by the adiabatic approximation,
the generating function is defined at each time and
the zeros oscillate as a function of time.
However, they do not represent the zeros of the global generating function.
It would be an interesting problem to clarify how
to describe the periodically driven systems
by the dynamical Lee--Yang zeros.

In this paper, we study distributions of the dynamical Lee--Yang zeros
by using the two-state classical master equation in several situations.
We show that the number of zeros is controllable by discretizing
the time variable.
Then, we can find a nontrivial distribution of zeros in the long-time limit.
To study periodically driven systems, 
we use the adiabatic approximation in the slow-driving
regime~\cite{Thouless, SN} and the Floquet--Magnus expansion 
in the fast-driving regime~\cite{Magnus, BCOR}.

The organization of the paper is as follows.
In Sec.~\ref{continuous}, we treat the continuous-time master equation
to reproduce known results on the dynamical Lee--Yang zeros.
The analysis is extended to the discrete-time master equation
to obtain a nontrivial distribution of zeros in Sec.~\ref{discrete}.
Then, we study a periodically driven system 
from an adiabatic picture in Sec.~\ref{periodic}.
We also study the continuous-time system with a high frequency
in Sec.~\ref{fm}.
Section \ref{summary} is devoted to summary.

\section{Continuous-time process}
\label{continuous}

\begin{figure}[t]
\begin{center}
\includegraphics[width=0.80\columnwidth]{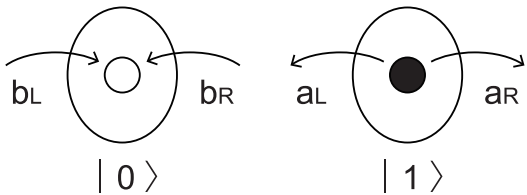}
\caption{Two-state classical stochastic processes.
The system takes two states $|0\rangle$ and $|1\rangle$.
Each of $a_{\rm L}$, $b_{\rm L}$, $a_{\rm R}$, and $b_{\rm R}$ represents
a transition rate.
}
\label{fig01}
\end{center}
\end{figure}

When we pursue an analogy between the current generating function
for stochastic processes
and the partition function in statistical mechanics,
it is reasonable to study the long-time behavior of the stochastic processes.
The time evolution of the probability distribution is generally given by
the master equation under the assumption that the process is Markovian.
To make the discussion concrete, we treat
the system depicted in Fig.~\ref{fig01}.
The system takes two states $|0\rangle=(1,0)^{\rm T}$ (``empty'')
and $|1\rangle=(0,1)^{\rm T}$ (``filled''),
and we observe flows among left and right leads.
When the transition-rate matrix in the master equation is time independent, 
the dynamical Lee-Yang zeros were obtained 
in previous studies~\cite{IA, FG, HFG, BMPGF}.
In this section, we summarize the basic definitions and notations, 
and reproduce the known results together with several small findings.

The time evolution of the probability distribution of states
$|p(t)\rangle=p_0(t)|0\rangle+p_1(t)|1\rangle$ is described by
the master equation.
We introduce the counting field $\chi$ to
measure flows between the system and the right lead
and define the modified master equation~\cite{SN}
\be
 \frac{\partial}{\partial t}|p^\chi(t)\rangle = w^\chi|p^\chi(t)\rangle,
\ee
with the transition-rate matrix
\be
 w^\chi
 =\left(\begin{array}{cc} -b & a^\chi \\ b^\chi & -a \end{array}\right)
 =\left(\begin{array}{cc} -(b_{\rm L}+b_{\rm R}) & a_{\rm L}+a_{\rm R}e^\chi \\
 b_{\rm L}+b_{\rm R}e^{-\chi} & -(a_{\rm L}+a_{\rm R}) \end{array}\right),
 \label{w}
\ee
where each of $a_{\rm L}$, $b_{\rm L}$, $a_{\rm R}$, and $b_{\rm R}$
represents a transition rate and takes a positive value.
The generating function is obtained from $|p^\chi(t)\rangle$ as
$\langle I|p^\chi(t)\rangle$ where $\langle I|=\sum_{n=0}^1\langle n|$.
Since the probabilistic properties of $|p(t)\rangle$ are maintained 
throughout the time evolution, 
the off-diagonal components of $w^{\chi=0}$ are non-negative and 
the diagonal components are determined from the relation 
$\langle I|w^{\chi=0}=0$.
The counting field is incorporated only in the off-diagonal components.
To describe the long-time limit, we define
the cumulant generating function
\be
 g(\chi)=\lim_{t\to\infty}\frac{1}{t}\ln \langle I|p^\chi(t)\rangle.
\ee
The $k$th moment of the current is given by
\be
 J_k=
 \left.\frac{\partial^k}{\partial \chi^k}g(\chi)\right|_{\chi=0}.
 \label{current}
\ee
The current generating function is obtained from 
the Legendre transformation of $g(\chi)$ as we demonstrate below.

The modified master equation is solved by using the spectral representation
of $w^\chi$ as
\be
 w^\chi=\sum_{n=0}^1\lambda_n^\chi |R_n^\chi\rangle\langle L_n^\chi|,
\ee
where $\lambda_n^\chi$ denotes the eigenvalue, and
$\langle L_n^\chi|$ and $|R_n^\chi\rangle$ are the corresponding
left and right eigenstates respectively.
The eigenstates satisfy the orthonormal relation 
$\langle L_m^\chi|R_n^\chi\rangle=\delta_{m,n}$ and
the resolution of unity $\sum_n|R_n^\chi\rangle\langle L_n^\chi|=1$.
Although the diagonalization of the general transition-rate matrix is 
not always possible, 
we can find the spectral representation in the present two-state case
as we see below.
The eigenvalues are explicitly written as
\be
 \lambda_{0,1}^\chi = -\frac{a+b}{2}
 \pm\frac{a+b}{2}\sqrt{\frac{(e^\chi-z_1)(e^\chi-z_2)}
 {e^\chi(1-z_1)(1-z_2)}}, \label{lambda}
\ee
where $z_1$ and $z_2$ are given by
\be
 z_{1,2} &=&
 -\frac{1}{2}\left[\frac{(a+b)^2}{4b_{\rm L}a_{\rm R}}
 -1-\frac{a_{\rm L}b_{\rm R}}{b_{\rm L}a_{\rm R}}\right] \nonumber\\
 &&\pm\sqrt{
 \frac{1}{4}\left[\frac{(a+b)^2}{4b_{\rm L}a_{\rm R}}
 -1-\frac{a_{\rm L}b_{\rm R}}{b_{\rm L}a_{\rm R}}\right]^2
 -\frac{a_{\rm L}b_{\rm R}}{b_{\rm L}a_{\rm R}}}. \label{z12}
\ee
The modified master equation for a given initial condition
$|p(0)\rangle$ is solved to give the generating function as
\be
 \langle I|p^\chi(t)\rangle =
 e^{\lambda_0^{\chi}t}\langle I|R_0^\chi\rangle\langle L_0^\chi|p(0)\rangle
 +e^{\lambda_1^{\chi}t}\langle I|R_1^\chi\rangle\langle L_1^\chi|p(0)\rangle.
 \nonumber\\
\ee
The component $n=0$ represents the stationary state since
$\lambda_0^\chi>\lambda_1^\chi$ holds for arbitrary real values of $\chi$.
We also note that
$\lambda_0^{\chi=0}=0$ and $\lambda_1^{\chi=0}=-(a+b)<0$.

We are interested in zeros of the generating function
by extending $\chi$ to complex values.
They are obtained by solving
\be
 \lambda_0^\chi-\lambda_1^\chi =
 \frac{1}{t}\ln\left(
 -\frac{\langle I|R_1^\chi\rangle\langle L_1^\chi|p(0)\rangle}
 {\langle I|R_0^\chi\rangle\langle L_0^\chi|p(0)\rangle}\right). \label{gap}
\ee
Although the general solution is dependent on $t$ and on
the initial condition, the equation reduces to
$\lambda_0^\chi=\lambda_1^\chi$ at large $t$.
From the explicit forms of the eigenvalues in Eq.~(\ref{lambda}),
we conclude that the dynamical Lee--Yang zeros
are given by $e^{\chi}=z_1$ and $e^{\chi}=z_2$.
These points are negative, which means that
the zeros are located on inaccessible region
in complex plane of $z=e^{\chi}$.
The cumulant generating function is given by
\be
 g(\chi)=\lambda_0^\chi
 = -\frac{a+b}{2}+\frac{a+b}{2}
 \sqrt{\frac{(e^{\chi}-z_1)(e^{\chi}-z_2)}{e^{\chi}(1-z_1)(1-z_2)}}.\nonumber\\
 \label{gchi}
\ee
This expression was obtained in previous works~\cite{IA, FG, HFG, BMPGF}.

The representation of the cumulant generating function by zeros
is convenient to find nontrivial relations of the current distributions.
We write $J_k$ in Eq.~(\ref{current}) as $J_k=(a+b)j_k/2$.
Then, $j_k$ is only dependent on $z_1$ and $z_2$.
For example, we have
\be
 j_1=\frac{1}{2}\left(\frac{1}{1-z_1}+\frac{1}{1-z_2}-1\right).
\ee
The higher-order moments are calculated in a similar way.
As they are parametrized only by $z_1$ and $z_2$,
we can find nontrivial relations between the moments.
Up to the fourth order of the cumulants, we obtain
\be
 && j_3+3j_2j_1-j_1=0, \\
 && j_4+3j_2^2-12j_2j_1^2-j_2+3j_1^2=0.
\ee
As far as we understand, these relations have not been obtained 
in previous studies.

The current distribution function $P(J,t)$ is calculated from
the Legendre transformation $\varphi(J)=g(\chi)-J\chi$ as
$P(J,t)\sim \exp(t\varphi(J))$ at large $t$.
The cumulant generating function in Eq.~(\ref{gchi})
is invariant under the transformation
$e^{\chi}\to z_1z_2e^{-\chi}$:
\be
 g(-\chi+\ln(z_1z_2))= g(\chi).
\ee
In the Legendre transformation,
the relation between $J$ and $\chi$ is given by
$J(\chi)=\partial_\chi g(\chi)$.
Then, $J(-\chi+\ln(z_1z_2))= -J(\chi)$ and we have
\be
 \varphi(J)-\varphi(-J)=J\ln\sqrt{z_1z_2}.
\ee
In the context of nonequilibrium thermodynamics,
this relation is known as the fluctuation theorem.
The product of the zeros $z_1z_2=a_{\rm L}b_{\rm R}/b_{\rm L}a_{\rm R}$
gives the affinity ${\cal A}=-\ln (z_1z_2)$ which represents
a bias between the left and right leads.

\section{Discrete-time process}
\label{discrete}

The polynomial representation of the generating function
in Eq.~(\ref{Zpoly}) is possible
only when the number of events is finite.
Equation (\ref{gchi}) does not have a simple polynomial form and
it is not clear why the number of zeros is given by 2.
We note that the number of zeros is not related to
the number of states in the master equation, 
as we confirm below.

Here we treat a discrete-time process.
It is suitable to describe rare events where
the charge transfer occurs sporadically.
We set a finite-time interval $\Delta t$ and
the total process time is discretized as $t=2M\Delta t$.
In this setting, the charge transfer is counted discretely and
the number of events is controllable by the integer parameter $M$.
In the following analysis, 
we take the long-time limit $M\to\infty$ while keeping
$\Delta t$ finite.

The discrete master equation is given by
\be
 |p(k)\rangle = (1+W)|p(k-1)\rangle,
\ee
where the integer index $k$ denotes a discretized time
$t/\Delta t$.
Each element of $1+W$ represents a probability
and must be smaller than unity.
The transition-rate matrix $w$ in the continuous limit is
obtained by using the relation $W=w\Delta t$.
When we include the counting field $\chi$ to $W$,
$W^\chi$ is a linear combination of $e^\chi$ and $e^{-\chi}$.
We write the transition matrix
\be
 W^\chi
 =\left(\begin{array}{cc} -B & A^\chi \\ B^\chi & -A \end{array}\right)
 =\left(\begin{array}{cc} -(B_{\rm L}+B_{\rm R}) & A_{\rm L}+A_{\rm R}e^\chi \\
 B_{\rm L}+B_{\rm R}e^{-\chi} & -(A_{\rm L}+A_{\rm R}) \end{array}\right). \nonumber\\
 \label{W}
\ee
Then, we see that the generating function 
$\langle I|p^\chi(2M)\rangle= \langle I|(1+W^\chi)^{2M}|p(0)\rangle$ 
can be factorized as 
\be
 \langle I|p^\chi(2M)\rangle
 = \prod_{k=0}^{M-1}
 \frac{(e^\chi-z_1(k))(e^\chi-z_2(k))}
 {e^\chi(1-z_1(k))(1-z_2(k))}. \label{z}
\ee
As in the continuous-time case,
we write $W^\chi$ as
\be
 W^{\chi}=\sum_{n=0}^1\Lambda_n^\chi|R_n^\chi\rangle\langle L_n^\chi|,
\ee
and the generating function
\be
 \langle I|p^\chi(2M)\rangle &=&
 (1+\Lambda_0^\chi)^{2M} \langle I|R_0^\chi\rangle
 \langle L_0^\chi|p(0)\rangle\nonumber\\
 && +(1+\Lambda_1^\chi)^{2M}\langle I|R_1^\chi\rangle
 \langle L_1^\chi|p(0)\rangle.
 \label{Zdiscrete}
\ee

We want to find the zeros of the generating function, 
$z_1(k)$ and $z_2(k)$ in Eq.~(\ref{z}).
They are obtained by solving $\langle I|p^\chi(2M)\rangle=0$.
In the continuous-time case, 
the solution is expressed as in Eq.~(\ref{gap}) and 
the right hand side of the equation is neglected in the large-time limit.
A similar analysis is applied for the present discrete-time case.
A notable difference in that case is that we need to introduce the factor 
$(e^{i\pi k/M})^{2M}=1$ with $k=0,1,\dots,M-1$.
We obtain the relation 
\be
 1+\Lambda_0^\chi =(1+\Lambda_1^\chi)e^{i\pi k/M}.
 \label{discretezero}
\ee
For a given $k$, there exist two zeros, $z_1(k)$ and $z_2(k)$.
They are obtained by solving the equation 
\be
 \sqrt{\frac{(z-z_1)(z-z_2)}{z(1-z_1)(1-z_2)}}
 =\frac{i}{R}\tan\frac{\pi k}{2M}, 
\ee
with respect to $z$.
Here, $z_{1,2}$ are given by Eq.~(\ref{z12}) with the replacement 
$(a_{\rm L}, b_{\rm L}, a_{\rm R}, b_{\rm R})\to
(A_{\rm L}, B_{\rm L}, A_{\rm R}, B_{\rm R})$, and 
$R=(A+B)/[2-(A+B)]$.
The explicit forms of $z_{1,2}(k)$ are given by 
\be
 && z_{1,2}(k)= \frac{z_1+z_2}{2}
 -\frac{(1-z_1)(1-z_2)\tan^2\frac{\pi k}{2M}}{2R^2} \nonumber\\
 &&
 \pm\sqrt{\left[\frac{z_1+z_2}{2}
 -\frac{(1-z_1)(1-z_2)\tan^2\frac{\pi k}{2M}}{2R^2}\right]^2-z_1z_2}. 
 \nonumber\\
\ee
The zeros are located on
the negative real axis of $z=e^\chi$ with
$z_1\le z_1(k)<0$ and $-\infty<z_2(k)\le z_2$.
We note that $z_{1}(0)=z_1$ and $z_{2}(0)=z_2$ represents the zeros
in the continuous-time case.
We also note that the relation
\be
 z_1(k)z_2(k)=z_1z_2=\frac{A_{\rm L}B_{\rm R}}{B_{\rm L}A_{\rm R}} \label{product}
\ee
holds for any $k$.
This property is due to the symmetry of the transition-rate matrix,
as discussed in the continuous-time case.

At the limit $\Delta t\to 0$, we obtain $R\to 0$ and
$(z_1(k), z_2(k))_{k\ne 0}\to (0,-\infty)$.
Contributions from $(z_1(k), z_2(k))$ with $k\ne 0$
do not affect the current since they satisfy
\be
 \frac{1}{1-z_1(k)}+\frac{1}{1-z_2(k)}-1
 =\frac{1-z_{1}z_{2}}{(1-z_1(k))(1-z_2(k))}\to 0 \nonumber\\
\ee
at the limit.
Then, the system is characterized by two zeros $z_1$ and $z_2$.
This behavior is consistent with the result in the previous section.
To find the nontrivial form of 
the cumulant generating function in Eq.~(\ref{gchi}),
we must carefully take the continuous-time limit $\Delta t\to 0$
and the long-time limit $M\to\infty$ simultaneously.

\begin{figure}[t]
\begin{center}
\includegraphics[width=0.60\columnwidth]{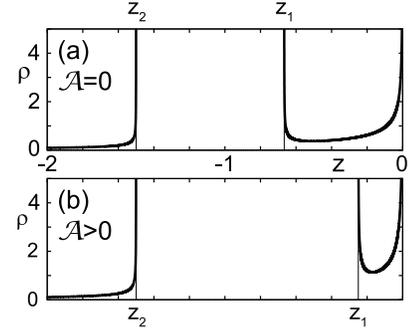}
\caption{The density of zeros in Eq.~(\ref{doz})
for discrete-time stochastic processes.
(a) We choose the parameters in 
the transition matrix in Eq.~(\ref{W}) as 
$(A_{\rm L}, B_{\rm L}, A_{\rm R}, B_{\rm R})=(0.3,0.2,0.3,0.2)$.
The affinity is given by ${\cal A}=-\ln (z_1z_2)=0$.
(b) $(A_{\rm L}, B_{\rm L}, A_{\rm R}, B_{\rm R})=(0.3,0.2,0.4,0.1)$ 
and ${\cal A}=-\ln (z_1z_2)=-\ln(3/8)>0$.
}
\label{fig02}
\end{center}
\end{figure}

We can find a continuous distribution of zeros
by taking the limit $M\to\infty$ while keeping $\Delta t$ finite.
At the limit, the density of zeros defined by 
\be
 \rho(z)=\frac{1}{2M}\sum_{k=0}^{M-1}\left(
 \delta (z-z_1(k))+\delta (z-z_2(k))\right)
\ee
is calculated from the number of $k$ within the interval $dz$.
We obtain 
\be
 \rho(z) &=& \left|
 \frac{-i}{2\pi}\frac{\partial}{\partial z}
 \ln\frac{1+\Lambda_0^{\chi=\ln z}}{1+\Lambda_1^{\chi=\ln z}}\right|
 \nonumber\\
 &=&
 \frac{1}{2\pi}\frac{R\sqrt{\frac{-(z-z_1)(z-z_2)}{z(1-z_1)(1-z_2)}}}{1+R^2
 \frac{-(z-z_1)(z-z_2)}{z(1-z_1)(1-z_2)}}
 \left|\frac{1}{z-z_1}+\frac{1}{z-z_2}-\frac{1}{z}\right|, \nonumber\\
 \label{doz}
\ee
for $z_1<z<0$ and $z<z_2$, and $\rho(z)=0$ otherwise.
This function is plotted in Fig.~\ref{fig02}.
The distribution of zeros is characterized by 
the edge points $z_1$ and $z_2$.
These points represent zeros in the continuous-time limit and 
the affinity is given by their product as ${\cal A}=-\ln (z_1z_2)$.

\begin{figure}[t]
\begin{center}
\includegraphics[width=0.80\columnwidth]{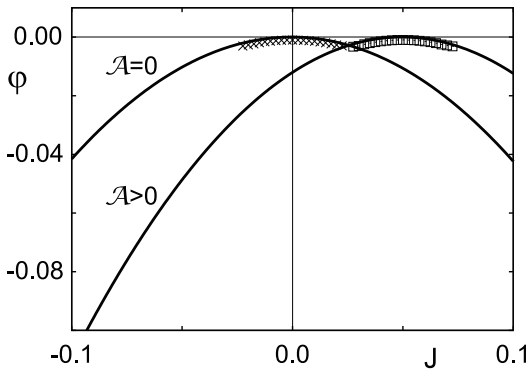}
\caption{
The solid lines denote the rate function $\varphi(J)$ in Eq.~(\ref{phi}) 
with the density of zeros in Fig.~\ref{fig02}.
We use the parametrization of $W$ in Fig.~\ref{fig02}(a)
for the line with ``${\cal A}=0$'', 
and the parametrization in Fig.~\ref{fig02}(b) for ``${\cal A}>0$''.
The marked points denote results from simulations.
We take $M=2000$ and the number of samples is $10^5$.
}
\label{fig03}
\end{center}
\end{figure}

The cumulant generating function is obtained from the density of zeros as
\be
 g(\chi) &=& \frac{1}{2M}\sum_{k=0}^{M-1}
 \ln\frac{(e^\chi-z_1(k))(e^\chi-z_2(k))}{e^\chi(1-z_1(k))(1-z_2(k))}
 \nonumber\\
 &=& \int dz\,\rho(z)\ln\frac{e^\chi-z}{1-z}
 -\frac{1}{2}\chi. \label{gen}
\ee
The current distribution function $P(J,2M)\sim\exp(2M\varphi(J))$
is also written by the density of zeros.
By using the Legendre transformation, we can write 
\be
 \varphi(J) = \int dz\,\rho(z)\ln\frac{e^\chi-z}{1-z}
 -\left(J+\frac{1}{2}\right)\chi, \label{phi}
\ee
where the relation between $J$ and $\chi$ is given by 
\be
 J=\int dz\,\rho(z)\frac{e^\chi}{e^\chi-z}-\frac{1}{2}. \label{J}
\ee
$\varphi(J)$ is plotted in Fig.~\ref{fig03}.


To confirm that the description by the zeros gives a reasonable result, 
we calculate the current distribution from numerical simulations.
The state of the system, $|0\rangle$ or $|1\rangle$, at each time $k$
is denoted by $n(k)=0$ or 1.
We generate a uniform random number $r(k)$ with $0\le r(k)\le 1$ 
at each $k$ to write the state $n(k)$ as 
\be
 n(k) &=& (1-n(k-1))\frac{1+{\rm sgn}(B-r(k))}{2} \nonumber\\
 && +n(k-1)\frac{1-{\rm sgn}(A-r(k))}{2}.
\ee
The net flow to the right lead is calculated as 
\be
 j(k)&=& j(k-1)-(1-n(k-1))\frac{1+{\rm sgn}(B_{\rm R}-r(k))}{2} \nonumber\\
 && +n(k-1)\frac{1+{\rm sgn}(A_{\rm R}-r(k))}{2},
\ee
with the initial condition $j(0)=0$.
Then, the current is given by $J=j(2M)/2M$.
We produce discrete-time sequences of $M=2000$ steps 
from a given set of transition probabilities.
The current distribution $P(J,2M)\sim\exp(2M\varphi(J))$ 
is obtained from $10^5$ samples.
The result is plotted in Fig.~\ref{fig03}.
We see that the result is consistent with that from the zeros.

\section{Periodically driven process}
\label{periodic}

When the transition probability fluctuates periodically, 
we observe a nontrivial distribution of the current. 
It is an interesting problem to study the properties of the system
by using the dynamical Lee--Yang zeros.
Although the complete study of the dynamical effects is a difficult 
problem in general, we can use the adiabatic approximation 
when the transition probability slowly changes as a function of time.
It is well known in the adiabatic regime that 
the nontrivial geometric effect is observed 
in the current distributions~\cite{Thouless, SN}.
In this section, we study the dynamical Lee--Yang zeros 
in periodically driven processes.
As we mentioned in Sec.~\ref{introduction}, 
the dynamical Lee--Yang zeros cannot be found in the continuous-time case 
when we use the adiabatic approximation.
This problem does not arise in the discrete-time case.
By extending the analysis of the previous section, 
we study periodically driven systems in the discrete-time case.

\subsection{Discrete-time formulation}

We denote the period of the oscillation by an integer $N$ and 
write the probability distribution as
\be
 |p^\chi(2MN)\rangle = (U_N^\chi)^{2M}|p(0)\rangle,
\ee
where
\be
 U_N^\chi = (1+W^\chi(N))(1+W^\chi(N-1))\cdots (1+W^\chi(1)). \nonumber\\
\ee
The transition matrix at discrete time $k=t/\Delta t$ is
denoted by $W(k)$ and satisfies $W(N+1)=W(1)$.
The total time is given by $t=2MN\Delta t$.
The generating function is written as in Eq.~(\ref{z}) 
with the replacement $M\to MN$.
The number of zeros is given by $2MN$.
When we use the spectral decomposition
\be
 U^\chi_N = \sum_{n=0}^1 (1+\Lambda_n^\chi)|R_n^\chi\rangle\langle L_n^\chi|,
\ee
the zeros of the generating function are given 
by solving Eq.~(\ref{discretezero}).

We consider the case where the transition matrix with $\chi$ 
is written in a form of Eq.~(\ref{W}) at each time. 
Then, the eigenvalues of $U_N^\chi$ are written as 
\be
 1+\Lambda_n^\chi 
 &=& \frac{1}{2}{\rm Tr}\, U_N^\chi
 \pm \left(1-\frac{1}{2}{\rm Tr}\, U_N^{\chi=0}\right)
 \nonumber\\
 &&\times\sqrt{\frac{(e^\chi-z_1)\cdots(e^\chi-z_{2N})}
 {(e^\chi)^N(1-z_1)\cdots(1-z_{2N})}}, \label{lambdan}
\ee
to define $\{z_1,z_2,\dots,z_{2N}\}$.
By using this form in Eq.~(\ref{discretezero}), 
we can find $2N$ zeros 
$\{z_1(k),z_2(k),\dots,z_{2N}(k)\}$ at each $k$.
The zeros are located in several separated domains where  
the function in the square root in Eq.~(\ref{lambdan}) is negative.
The number of the domains is given by $2\times \lceil{N/2}\rceil$.
Each point $z_i(0)=z_i$ ($i=1,2,\dots,2N$) denotes an edge of a domain.

The density of zeros in the present case is defined as 
\be
 \rho(z)
 = \frac{1}{2MN}\sum_{k=0}^{M-1}\sum_{i=1}^{2N}\delta (z-z_i(k)).
\ee
When we assume that all of $\{z_1,z_2,\dots,z_{2N}\}$ 
take real (negative) values, 
by taking the limit $M\to\infty$, we can write 
\be
 \rho(z)
 =\left|\frac{-i}{2\pi N}\frac{\partial}{\partial z}
 \ln\frac{1+\Lambda_0^{\chi=\ln z}}{1+\Lambda_1^{\chi=\ln z}}\right| 
\ee
in the domains of definition.
The cumulant generating function is calculated from 
the second line of Eq.~(\ref{gen}).

\subsection{Dynamical current and geometrical current}

We examine the relation between 
the current distribution and the density of zeros.
Before studying the discrete-time system, we summarize the result 
in the continuous-time system~\cite{SN, TFHH, THFH}.
When the system is driven periodically, 
each of $a_{\rm L}$, $b_{\rm L}$, $a_{\rm R}$, and $b_{\rm R}$
is represented as a function of $\theta=\omega t$ with 
the period $2\pi$.
The current consists of the dynamical part and the geometrical part: 
$J=J_{\rm d}+J_{\rm g}$.
The dynamical current is written by using the largest eigenvalue of 
the transition-rate matrix as
\be
 J_{\rm d}
 &=& \frac{\partial}{\partial\chi}
 \left.\int_0^{2\pi}\frac{d\theta}{2\pi}\,\lambda_0^\chi(\theta)\right|_{\chi=0}
 \nonumber\\
 &=& \int_0^{2\pi}\frac{d\theta}{2\pi}\,
 \frac{b_{\rm L}(\theta)a_{\rm R}(\theta)-a_{\rm L}(\theta)b_{\rm R}(\theta)}
 {a(\theta)+b(\theta)}.
\ee
This is independent of the frequency $\omega$. 
The geometrical part is written by using 
the expansion with respect to $\omega$.
At the first order, the geometrical current is written by using 
the Berry curvature as 
\be
 J_{\rm g}
 &=& -\omega\frac{\partial}{\partial\chi}
 \left.\int_0^{2\pi}\frac{d\theta}{2\pi}\,
 \langle L_0^\chi(\theta)|\partial_\theta R_0^\chi(\theta)\rangle
 \right|_{\chi=0}  \nonumber\\
 &=& \omega\int_0^{2\pi}\frac{d\theta}{2\pi}\,
 \frac{a_{\rm R}(\theta)+b_{\rm R}(\theta)}{a(\theta)+b(\theta)}\frac{d}{d\theta} 
 \frac{a(\theta)}{a(\theta)+b(\theta)}.
\ee
The current is represented by a flux penetrating a surface 
in parameter space~\cite{SN, Berry}.

To examine the geometric property of the system, 
it is useful to write each parameter as follows:
\be
 && \frac{a_{\rm L}(\theta)}{a+b}
 =\frac{1+r(\theta)}{2}\sin^2\left(\frac{\phi_a(\theta)}{2}\right), 
 \label{p1}\\
 && \frac{a_{\rm R}(\theta)}{a+b}
 =\frac{1+r(\theta)}{2}\cos^2\left(\frac{\phi_a(\theta)}{2}\right), \\
 && \frac{b_{\rm L}(\theta)}{a+b}
 =\frac{1-r(\theta)}{2}\sin^2\left(\frac{\phi_b(\theta)}{2}\right), \\
 && \frac{b_{\rm R}(\theta)}{a+b}
 =\frac{1-r(\theta)}{2}\cos^2\left(\frac{\phi_b(\theta)}{2}\right).
 \label{p4}
\ee
$\phi_a$ and $\phi_b$ are real and $r$ satisfies $-1\le r(\theta)\le 1$.
Since $a+b$ determines the overall time scale, we take it constant.
Then, the dynamical current is written as 
\be
 J_{\rm d}
 = \frac{a+b}{8}\int_0^{2\pi}\frac{d\theta}{2\pi}\,(1-r^2(\theta))
 (\cos\phi_a(\theta)-\cos\phi_b(\theta)). \nonumber\\
\ee
It is nonzero when $\phi_a\ne\phi_b$ as we can understand from the expression
of the affinity at each time: 
\be
 {\cal A}(\theta)=\ln
 \frac{b_{\rm L}(\theta)a_{\rm R}(\theta)}{a_{\rm L}(\theta)b_{\rm R}(\theta)}
 =\ln\left(\frac{\tan\frac{\phi_b(\theta)}{2}}
 {\tan\frac{\phi_a(\theta)}{2}}\right)^2.
\ee
On the other hand, the geometrical current is given by 
\be
 J_{\rm g}
 &=& \omega\int_0^{2\pi}\frac{d\theta}{2\pi}\,
 \frac{1}{8}\frac{d r(\theta)}{d\theta} \nonumber\\
 &&\times
 \left[
 (1+r(\theta))\cos\phi_a(\theta)+(1-r(\theta))\cos\phi_b(\theta)
\right]. \nonumber\\ \label{jg}
\ee
To find a nonzero value of $J_{\rm g}$, we need that 
$r$ and, at least, one of $\phi_a$ and $\phi_b$ has
a nontrivial time dependence.

In a typical parametrization, the dynamical part is the dominant 
contribution to the current. 
The dynamical current is calculated from the instantaneous eigenvalue 
of the transition-rate matrix.
The corresponding approximation in the discrete-time case
is given by 
\be
 g(\chi)=\frac{1}{N}\ln(1+\Lambda_0^\chi)
 \sim \frac{1}{N}\sum_{k=1}^N \ln(1+\Lambda_0^\chi(k)), \label{ad0}
\ee
where $\Lambda_0^\chi(k)$ denotes the largest eigenvalue of $W^\chi(k)$.
The geometrical current is typically a small quantity and 
becomes important only when the dynamical current is absent.

\begin{figure}[t]
\begin{center}
\includegraphics[width=0.80\columnwidth]{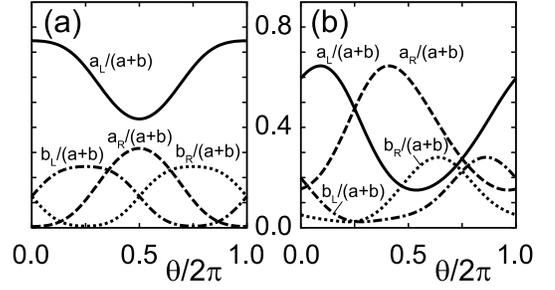}
\caption{Two protocols used in Sec.~\ref{periodic}.
$J_{\rm d}\ne 0$ and $J_{\rm g}=0$ for the protocol in the panel (a), and
$J_{\rm d}= 0$ and $J_{\rm g}\ne 0$ in the panel (b).
}
\label{fig04}
\end{center}
\end{figure}

The present parametrization allows us to study each part of the current 
separately.
In the first protocol, we use 
\be
 && r(\theta)=\frac{1}{2}, \\
 && \phi_a(\theta)=\frac{3\pi}{4}+\frac{\pi}{5}\cos\theta, \\
 && \phi_b(\theta)=\frac{\pi}{2}+\frac{2\pi}{5}\sin\theta.
\ee
The corresponding behavior of parameters 
$a_{\rm L}$, $b_{\rm L}$, $a_{\rm R}$, and $b_{\rm R}$ is shown in 
Fig.~\ref{fig04}(a).
In this case, $J_{\rm g}=0$ in the continuous-time case 
and the dominant contribution of the current 
comes from the dynamical part.

\begin{figure}[t]
\begin{center}
\includegraphics[width=0.80\columnwidth]{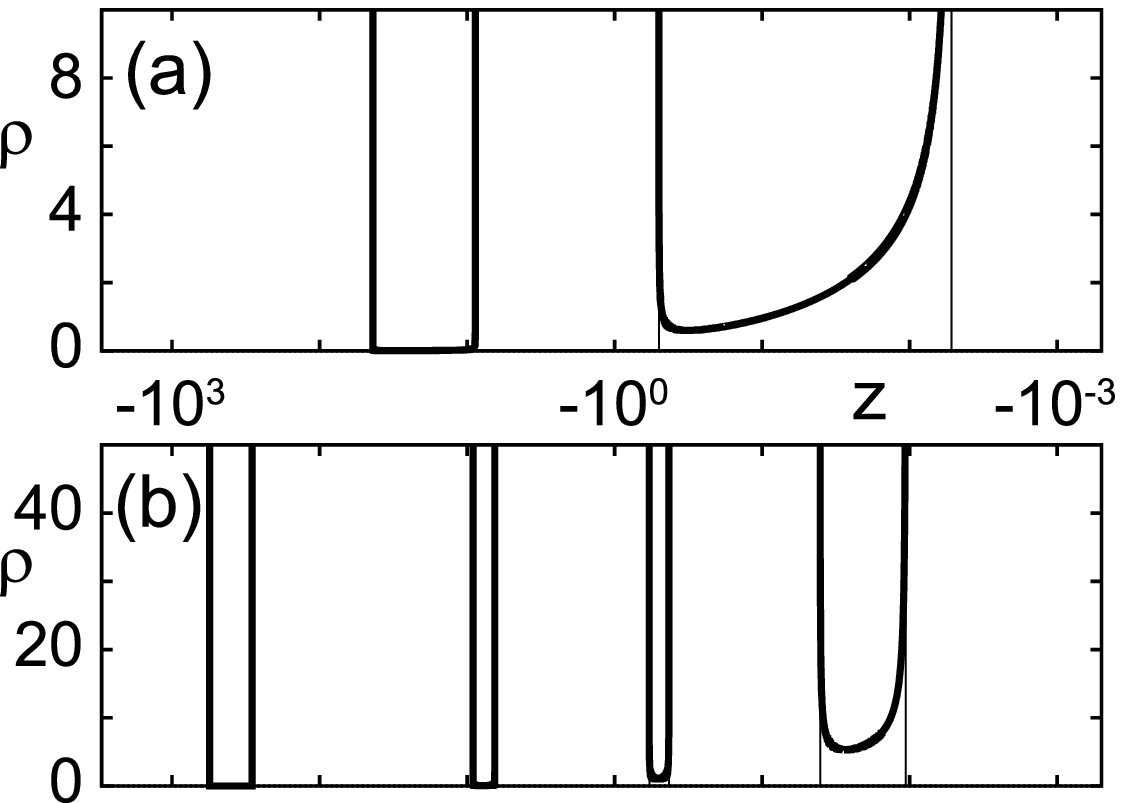}
\caption{The density of zeros in the first protocol of Fig.~\ref{fig04}(a) 
where $J_{\rm d}\ne 0$ and $J_{\rm g} = 0$ 
in the corresponding continuous-time process. 
We set $N=2$ in the panel (a) and $N=4$ in the panel (b).
}
\label{fig05}
\end{center}
\end{figure}
\begin{figure}[t]
\begin{center}
\includegraphics[width=0.80\columnwidth]{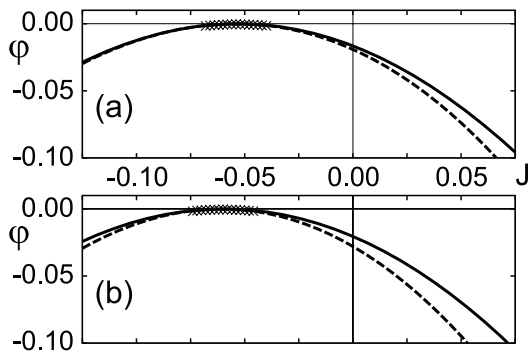}
\caption{$\varphi(J)$ in the first protocol.
We set $N=2$ in the panel (a) and $N=4$ in the panel (b).
In each panel, 
the solid line denotes the result calculated from Eq.~(\ref{phi}).
We use the density of zeros in Fig.~\ref{fig05}.
The dashed line denotes the approximation in Eq.~(\ref{ad0}).
The marked points denote simulation results.
We set $MN = 5000$ and the number of samples to $10^8$ in the simulations.
}
\label{fig06}
\end{center}
\end{figure}

In the discrete-time case, we consider the process by using 
the discretized protocol.
In Fig.~\ref{fig05}, we plot the density of zeros at $N=2$ and $N=4$.
We find that each $z_i$ ($i=1,\dots,2N$) in these cases takes  
a negative value.
Then, the zeros continuously distribute 
on the negative real axis of $z$ with 
$z_{i} \ge z\ge z_{i+1}$ ($i=1,2,\dots,2N-1$) 
where we set $0>z_1>z_2>\cdots>z_{2N}$.

The corresponding rate function $\varphi(J)$, 
calculated from Eq.~(\ref{phi}), is shown in Fig.~\ref{fig06}.
We compare three results, $\varphi(J)$ exactly calculated from $\rho(z)$,
$\varphi(J)$ from the approximation in Eq.~(\ref{ad0}), 
and $\varphi(J)$ from the simulations.
As we see in Fig.~\ref{fig06}, 
the result from zeros is consistent with that from the simulations.
We also observe that the average current is well described 
by the approximation in Eq.~(\ref{ad0}).

We next consider the second protocol: 
\be
 && r(\theta)=\frac{1}{2}+\frac{2}{5}\sin\theta, \\
 && \phi_a(\theta)=\phi_b(\theta)=\frac{\pi}{2}+\frac{\pi}{5}\cos\theta.
\ee
The corresponding behavior of parameters 
$a_{\rm L}$, $b_{\rm L}$, $a_{\rm R}$, and $b_{\rm R}$ is shown in 
Fig.~\ref{fig04}(b).
In this case, the current in the continuous-time case is purely geometric.
${\cal A(\theta)}=0$ at each time, which means that 
the instantaneous dynamical current is exactly equal to zero.
On the other hand, the flux in Eq.~(\ref{jg}) is nonzero and 
we can observe a nonzero geometrical current.

\begin{figure}[t]
\begin{center}
\includegraphics[width=0.80\columnwidth]{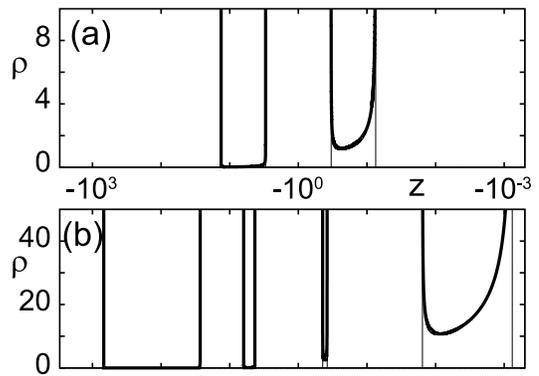}
\caption{
The density of zeros in the second protocol 
of Fig.~\ref{fig04}(b) where 
$J_{\rm d}= 0$ and $J_{\rm g}\ne 0$ in the corresponding continuous-time process. 
We set $N=2$ in the panel (a) and $N=4$ in the panel (b).
}
\label{fig07}
\end{center}
\end{figure}
\begin{figure}[t]
\begin{center}
\includegraphics[width=0.80\columnwidth]{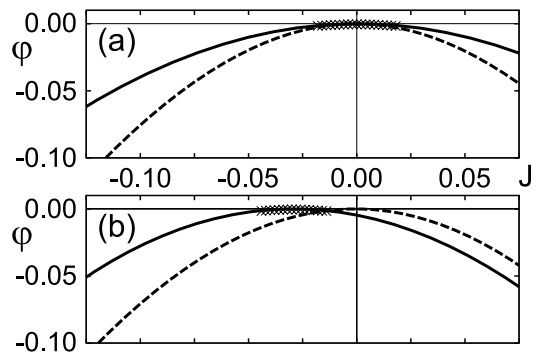}
\caption{
$\varphi(J)$ in the second protocol.
We use the density of zeros in Fig.~\ref{fig07}.
See the caption of Fig.~\ref{fig06} for other remarks.
}
\label{fig08}
\end{center}
\end{figure}

We study how the geometric effect appears in the discrete-time case.
At $N=2$ and $N=4$, 
each $z_i$ ($i=1,\dots,2N$) takes a negative value also in this case.
The results of $\rho(z)$ are shown in Fig.~\ref{fig07}
and $\varphi(J)$ in Fig.~\ref{fig08}.

In the case of $N=2$, the current distribution is almost symmetric and 
the magnitude of the average current takes a small value.
$\varphi(J)$ at small $|J|$ is well described by the approximation 
in Eq.~(\ref{ad0}).
At $N=2$, the protocol trajectory in parameter space 
does not enclose any surface and no geometric effect is observed.
This result is significantly changed when we consider the $N=4$ case.
The approximation in Eq.~(\ref{ad0}) does not give any reasonable result 
and we observe a nonzero average current.
We find that simulation results are well fitted by the result from zeros.
Since the condition of the adiabatic approximation is not obvious 
in the discrete-time case, 
it is remarkable to find that the geometric effect can be seen 
at small $N$.

\section{Floquet--Magnus expansion}
\label{fm}

We have described the systems with small frequency $\omega$ 
by using the adiabatic approximation in the previous section.
A similar systematic analysis is possible 
when $\omega$ takes a large value.
When the transition-rate matrix drastically changes in short times, 
it is not appropriate to treat the discrete-time system and 
we consider the continuous-time case in this section.

In contrast to the adiabatic case, 
we can easily find two zeros in two-state systems and study 
the fluctuation effects on the current distributions.
When the period of the modulation is very small,
we can expect that the system is basically described by
the averaged transition-rate matrix
$\bar{w}^\chi\sim \bar{w}_0^\chi =
\int_0^{2\pi}\frac{d\theta}{2\pi} w^\chi(\theta)$.
The correction can be expressed by using a series expansion
of the inverse frequency,
which is known as the Floquet--Magnus expansion~\cite{Magnus, BCOR}.

We define the effective transition-rate matrix $\bar{w}^\chi$
from the relation
\be
 \exp\left(\frac{2\pi}{\omega}\bar{w}^\chi\right)
 ={\rm T}\exp\left(\int_0^{2\pi/\omega} dt\,w^\chi(\omega t)\right),
\ee
where T denotes that the time-evolution operator is represented by
the time-ordered product.
In the Floquet--Magnus expansion, this effective matrix is represented as
$\bar{w}^\chi=\sum_{k=0}^\infty \bar{w}_k^\chi$.
$\bar{w}_k^\chi$ is proportional to $1/\omega^k$ and
the explicit forms at $k=1$ and 2 are given as follows:
\be
 \bar{w}_1^\chi= \frac{1}{2\omega}
 \int_0^{2\pi}\frac{d\theta_1}{2\pi}\int_0^{\theta_1}d\theta_2
 [w^\chi(\theta_1),w^\chi(\theta_2)],
\ee
\be
 \bar{w}_2^\chi &=& \frac{1}{6\omega^2}
 \int_0^{2\pi}\frac{d\theta_1}{2\pi}\int_0^{\theta_1}d\theta_2
 \int_0^{\theta_2}d\theta_3
 \nonumber\\
 && ([w^\chi(\theta_1),[w^\chi(\theta_2),w^\chi(\theta_3)]\nonumber\\
 && +[w^\chi(\theta_3),[w^\chi(\theta_2),w^\chi(\theta_1)]).
\ee

To give a concrete discussion, we treat the two-state case.
Since the expansion keeps the trace of the transition-rate matrix invariant,
we can generally write
\be
 \bar{w}^\chi=-\frac{\bar{a}+\bar{b}}{2}
 +\left(\begin{array}{cc}
 \alpha^\chi & \beta^\chi \\
 \gamma^\chi & -\alpha^\chi \end{array}\right),
\ee
where
\be
 \bar{a}=\int_0^{2\pi}\frac{d\theta}{2\pi} a(\theta),
\ee
and $\bar{b}$ is defined in a similar way.
The cumulant generating function is given by
the largest eigenvalue of $\bar{w}^\chi$ as 
\be
 g(\chi)=-\frac{\bar{a}+\bar{b}}{2}
 +\sqrt{(\alpha^\chi)^2+\beta^\chi\gamma^\chi}.
\ee
Up to the first order of the expansion,
each element is given respectively by
\be
 \alpha^\chi
 &\sim& \frac{\bar{a}-\bar{b}}{2}
 +\frac{1}{2\omega}\int_0^{2\pi}\frac{d\theta_1}{2\pi}
 \int_0^{\theta_1}d\theta_2\left(
 a^\chi(\theta_1)b^\chi(\theta_2)\right.\nonumber\\
 && \left. -b^\chi(\theta_1)a^\chi(\theta_2)\right),
\ee
\be
 \beta^\chi
 &\sim& \bar{a}^\chi
 +\frac{1}{2\omega}\int_0^{2\pi}\frac{d\theta_1}{2\pi}
 \int_0^{\theta_1}d\theta_2[
 (a(\theta_1)-b(\theta_1))a^\chi(\theta_2)
\nonumber\\ &&
 -a^\chi(\theta_1)(a(\theta_2)-b(\theta_2))],
\ee
\be
 \gamma^\chi
 &\sim& \bar{b}^\chi
 +\frac{1}{2\omega}\int_0^{2\pi}\frac{d\theta_1}{2\pi}
 \int_0^{\theta_1}d\theta_2[
 -(a(\theta_1)-b(\theta_1))b^\chi(\theta_2)
\nonumber\\ &&
 +b^\chi(\theta_1)(a(\theta_2)-b(\theta_2))].
\ee
From these relations, we can find that
the cumulant generating function takes a form
\be
 g(\chi)\sim
 -\frac{\bar{a}+\bar{b}}{2}
 +\frac{\bar{a}+\bar{b}}{2}
 \sqrt{\frac{(e^\chi-z_1)(e^\chi-z_2)}{e^\chi(1-z_1)(1-z_2)}}.
\ee
As in the zeroth order, the function is characterized by two zeros.

We can also examine the higher-order corrections.
Although it is a difficult task to find the explicit forms,
the expansion is systematically represented by multiple commutators.
We find that up to the $2n$th order of the expansion
\be
 g(\chi)\sim
 -\frac{\bar{a}+\bar{b}}{2}
 +\frac{\bar{a}+\bar{b}}{2}
 \sqrt{\prod_{k=0}^{n-1}\frac{(e^\chi-z_1(k))(e^\chi-z_2(k))}
 {e^\chi(1-z_1(k))(1-z_2(k))}}.\nonumber\\
\ee
Then, the average current is given by
\be
 J\sim \frac{\bar{a}+\bar{b}}{2}\sum_{k=0}^{n-1}
 \frac{1}{2}\left(\frac{1}{1-z_1(k)}+\frac{1}{1-z_2(k)}-1\right).
\ee
The current is generally characterized by many zeros
and the number of the zeros increases
as we increase the order of the expansion.

\begin{figure}[t]
\begin{center}
\includegraphics[width=0.80\columnwidth]{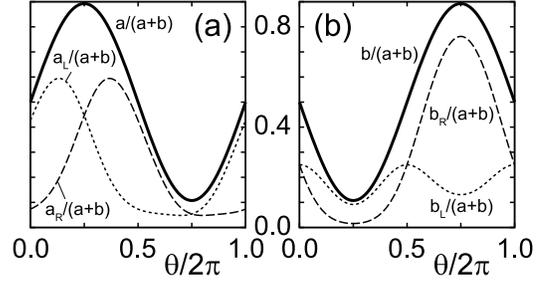}
\caption{Protocol used in Sec.~\ref{fm}.
(a) $a_{\rm L}$, $a_{\rm R}$, and $a=a_{\rm L}+a_{\rm R}$.
(b) $b_{\rm L}$, $b_{\rm R}$, and $b=b_{\rm L}+b_{\rm R}$.
}
\label{fig09}
\end{center}
\end{figure}
\begin{figure}[t]
\begin{center}
\includegraphics[width=0.80\columnwidth]{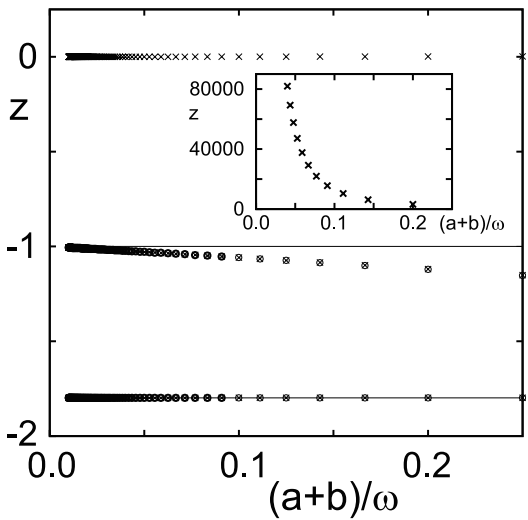}
\caption{Zeros obtained by the Floquet--Magnus expansion.
At the first-order expansion, we have two zeros at each frequency $\omega$,
which are denoted by points $\circ$.
At the second order, we have four zeros denoted by points $\times$
and one of them is shown in the inset.
}
\label{fig10}
\end{center}
\end{figure}
\begin{figure}[t]
\begin{center}
\includegraphics[width=0.80\columnwidth]{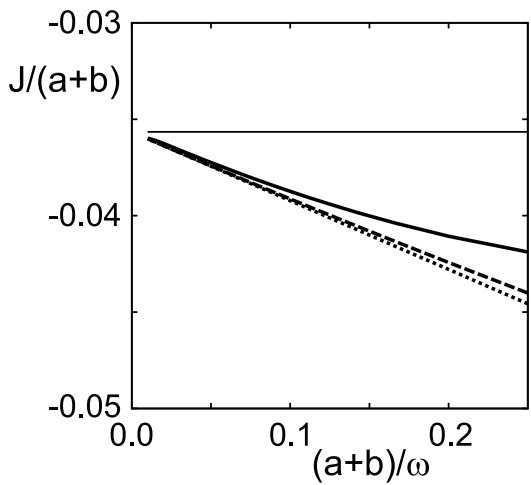}
\caption{Average current as a function of the inverse frequency.
The bold solid line denotes the exact result, dotted line
denotes the first-order accuracy of the Floquet--Magnus expansion,
dashed line denotes the second order of the expansion.}
\label{fig11}
\end{center}
\end{figure}

We show the result of the expansion up to the second order.
We set the protocol in Eqs.~(\ref{p1})-(\ref{p4}) with
\be
 && r=\frac{\pi}{4}\sin\theta, \\
 && \phi_a=\frac{\pi}{2}+\frac{\pi}{4}\cos\theta, \\
 && \phi_b=\frac{\pi}{2}+\frac{\pi}{4}\sin\theta,
\ee
as shown in Fig.~\ref{fig09}.
The zeros are plotted in Fig.~\ref{fig10} and
the corresponding current is in Fig.~\ref{fig11}.
We see that the current is properly described by using the zeros.
At the zeroth order of the expansion, two zeros
are located on the negative real axis.
They change as functions of $1/\omega$ by taking into account 
the higher-order corrections.
At the same time, when we increase the order of the expansion,
different zeros appear around the origin and
on the real axis far from the origin.
These zeros are treated in pairs and
give small corrections to the current as
\be
 \frac{1}{1-z_1(k)}+\frac{1}{1-z_2(k)}-1 \to 0,
\ee
at $|z_1(k)|\to \infty$ and $z_2(k)\to 0$.

To study the meaning of zeros,
we examine how the fluctuation theorem is affected by the expansion.
The rate is calculated from the Legendre transformation
$\varphi(J)=g(\chi)-J\chi$.
In the expansion up to $2n$th order,
the relation between $\chi$ and $J$ is obtained from
\be
 J(\chi) &\sim&
 \frac{\bar{a}+\bar{b}}{2}\sum_{k=0}^{n-1}
 \frac{1}{2}\left(\frac{1}{e^\chi-z_1(k)}+\frac{1}{e^\chi-z_2(k)}
 -\frac{1}{e^\chi}\right) \nonumber\\
 && \times
 \sqrt{\prod_{k=0}^{n-1}\frac{(e^\chi-z_1(k))(e^\chi-z_2(k))}
 {e^\chi(1-z_1(k))(1-z_2(k))}}.
\ee
$\chi$ with $e^\chi>0$ is uniquely determined for a given $J$.
Since it is difficult to obtain a compact form of the general solution,
we discuss the cases where $J$ takes extremal values.

When the absolute value of $J$ is large, $|J|\gg\bar{a}+\bar{b}$,
the corresponding counting field is also large $|\chi|\gg 1$ and
we can evaluate
\be
 J(\chi)\sim 
 \frac{\bar{a}+\bar{b}}{2} n(e^\chi)^{n/2}
 \sqrt{\prod_{k=0}^{n-1}\frac{1}{(1-z_1(k))(1-z_2(k))}}
 \nonumber\\
\ee
for $e^\chi\gg 1$ and 
\be
 J(\chi)\sim 
 -\frac{\bar{a}+\bar{b}}{2} n(e^\chi)^{-n/2}
 \sqrt{\prod_{k=0}^{n-1}\frac{z_1(k)z_2(k)}{(1-z_1(k))(1-z_2(k))}} 
 \nonumber\\
\ee
for $e^\chi\ll 1$.
Then, after some calculations we obtain
\be
 \varphi(J)-\varphi(-J) 
 \sim J\ln\left(\prod_{k=0}^{n-1} z_1(k)z_2(k)\right)^{1/n}
 \label{ft}
\ee
for $|J|\gg \bar{a}+\bar{b}$.
This relation implies that the affinity is given
by the geometric mean of zeros.
As we find in the above example,
the product $z_1(k)z_2(k)$ takes a finite value.

In the opposite limit at $J=0$,
the corresponding $e^\chi$ is obtained by solving
\be
 \sum_{k=0}^{n-1}
 \frac{(e^{\chi^*})^2-z_1(k)z_2(k)}{(e^{\chi^*}-z_1(k))(e^{\chi^*}-z_2(k))}=0.
\ee
Then, $\varphi(J)$ is obtained as $\varphi(J)\sim J\ln e^{\chi^*}$ and
\be
 \varphi(J)- \varphi(-J) \sim J\ln (e^{\chi^*})^2, 
\ee
for $|J|\ll \bar{a}+\bar{b}$.

We numerically find that the product $z_1(k)z_2(k)$ does not satisfy
Eq.~(\ref{product}) and is dependent on the index $k$,
which means that Eq.~(\ref{ft}) does not hold in the time-dependent system.
The fluctuation theorem still holds in that case,
but its consequence is not represented in a simple form
as in Eq.~(\ref{ft})~\cite{THFH}.

\section{Summary}
\label{summary}

We have studied dynamical Lee--Yang zeros in stochastic processes.
As we mentioned in the Introduction, 
the distribution of zeros has complete information about 
the statistical properties of the processes.
By discretizing the time variable, 
we can explicitly confirm that 
the number of zeros is related not to the number of states 
but to the number of events.
We find nontrivial distributions of zeros 
even in simple two-state processes.
The distribution of zeros is related to the current distributions 
and we expect that our discrete-time result   
can be confirmed experimentally by measuring processes with rare events.

Since our analysis is restricted to simple two-state systems, 
the zeros distribute only on the negative real axis of $z=e^{\chi}$.
In statistical mechanics, thermal phase transitions can be found  
by studying the distribution of zeros.
The locations of zeros represent the boundary of different phases 
in complex plane.
We have not found closed boundaries in the present analysis.
It may be an interesting problem to study more complex systems 
such as quantum dot systems coupled with external reservoirs
treated in previous studies~\cite{UEUA, SMY17}.

One of the important findings in the present study is that 
the method can be applied in the case of 
the periodically oscillating transition rate.
In the adiabatic case with slowly varying transition-rate matrix, 
the property of the density of zeros is   
dependent on the geometric properties of the protocols.
We also find in the fast-driving regime of the continuous-time case 
that the number of zeros depends on the order of the expansion.
These results show that it is important to study  
global distributions of zeros 
to characterize the dynamically fluctuating systems
instead of using a few number of zeros.
The distribution also reflects a fundamental symmetry of the system.
We expect that the principle of nonequilibrium thermodynamics can be 
studied along with the method developed in the present study.

\section*{Acknowledgments}
We thank Y.~Utsumi for valuable comments.
K.T. was supported by JSPS KAKENHI Grants No. JP20K03781 and No. JP20H01827.


\end{document}